\documentclass[12pt,a4paper]{article}

\title{\large\bf The test particle motion equations metrical form \\
          in a potential field}
\author{\large K.B. Korotchenko}
\date{\normalsize Russian Physics Journal, Vol. ??, No. 12, 1983}
\begin{document}
\maketitle
\begin{abstract}
     It is shown in the present work that the
     three-dimensional trajectories of an electrical test
     particle in potential fields may be regarded
     as geodesic lines lying on isotropic surfaces of some
     four-dimensional configurational space, the
     connection of which has tortion, while the
     transference is nonmetric.
\end{abstract}

     The starting point in the present work is the
     Einstein concept that the geometrization of an
     interaction consists in finding a metric space in
     which the test particles trajectories are
     geodesic lines \cite{p1}.

     An interesting method of metrization of arbitrary
     force interactions corresponding (to a definite
     extent) to this concept was proposed in \cite{p2}.  In this
     method of metrization, the test particles move
     along geodesic lines.  However, the force fields are related
     with the components of the connection tortion tensor
     of a pseudo-Euclidean space.  In this sense, the
     metrization of force interactions proposed in \cite{p2} does
     not correspond to the Einstein program, since the
     metric properties of the space in which the force
     fields act do not depend on these force fields.

     However, it is also of interest to consider a metric
     formulation of force interactions in which, as in \cite{p2},
     the test particles motion equations
     represent a special form of Newton's second law in
     four-dimensional form (constituting here the geodesic
     equation of some four-dimensional space $^kV_4 $)
     but the metric tensor and physical fields are interdependent.

     The construction of such a special geometric
     formulation of force interactions is the subject of
     the present work.  To avoid problems associated with
     the distinction between the concepts of a reference
     frame and a coordinate system \cite{p2}, different observers
     (i.e., reference frames) will be posed in
     accordance, generally speaking, with different
     four-dimensional spaces $^kV_4 $.

     To simplify the mathematical formalism, the
     test particles motion only in potential fields is
     considered in the present work.

     1.  The states of the test particles (of mass $m $
     and charge $e $) in the potential fields will be called
     classical states.  Correspondingly, all the
     characteristics of the particle describing its
     behavior in the classical state (trajectory, velocity,
     momentum, energy, etc.) will be called classical.

     It should be emphasized that below all classical
     characteristics are assumed to be specified with
     respect to one definite reference frame, which may be
     chosen in the form of any inertial frame (IF).

     2.  Suppose that $^kV_4 $ is a four-dimensional space with
     the metric
\begin{equation}
\label{e1}
 ^{(k)}dS^2\: =\: {^{(k)}}g_{oo}(x^i, t) c^2 dt^2\: +
               \: g_{ik} dx^i dx^k\: ,
\end{equation}
     where ($- g_{ik} $) is the metric tensor of the space $V_3 $
     (the "spatial" component of the Minkowski four-space $V_4 $).

     It is clear that any classical trajectory
     $x^i = x^i (t) $ may be regarded as a line
     on the isotropic surface $^kG_{o3} \subset {^k}V_4 $,
     defined by the equation
     ${^{(k)}}g_{oo}c^2 dt^2 = -þ g_{ik} dx^i dx^k $ , under the
     condition that along this line
\begin{equation}
\label{e2}
 {^{(k)}}g_{oo}(x^i (t), t)\: =\: v^i v_i/c^2\: ,
\end{equation}
     where $v^i $ is the particle velocity measured from the
     specified IF ($v_i = þ- g_{ik} v^k $).

     Thus, each point $p $ of the classical particle
     trajectory $x^i = x^i (t) $ in $V_3 $ may be regarded
     as lying on the isotropic surface
     $^kG_{o3} \subset {^k}V_4 $.

     It is understood that in $V_3 $ classical states of the
     particle are possible such that, at any point $p \in V_3 $,
     the particle velocity is independent on the trajectory
     along which the particle reaches this point (of
     course, with certain specified initial parameters).
     For such states, each point $p \in V_3 $ may also be
     regarded as a point of the isotropic surface
     $^kG_{o3} $ in $^kV_4 $.
     That is an isotropic surface
     $^kG_{o3} \subset {^k}V_4 $ may be
     constructed at such points of space $V_3 $. By
     changing the values of the initial parameters, a set
     of surfaces $^kG_{o3} $ covering the whole of
     $^kV_4 $ may be obtained. That is an imbedding
     \cite{p1} þ- enclosure in a space of higher
     dimensionality þ- may be constructed.

     3. According to Eq.(\ref{e1}), the method of enclosure
     described in Sec.2 should have the distinctive
     property that the geometry of the enclosing space
     $^kV_4 $
     should have no influence on the geometric properties
     of the enclosed space $V_3 $ (should not change
     the metric tensor $g_{ik} $). In other words, the imbedding
     must occur at those points of $^kV_4 $ at which the
     external curvature of the enclosed surface is zero.

     Then it follows from the GaussþVaingarten equations þ-
     see \cite{p3}, for example þ- that, with this imbedding, the
     absolute differential of the space $^kV_4 $ (denoted by
     ${^{(k)}}\nabla (\cdots) $) is defined by the equation
\begin{equation}
\label{e3}
 ^{(k)}\nabla A^{\mu}\: =\: ({^{(3)}}\nabla_i A^{\mu}) dx^i\: +
                       \: ({^{(4)}}\nabla_o A^{\mu}) dx^o\: .
\end{equation}
     Equation (\ref{e3}) may also be rewritten in the form
\begin{equation}
\label{e4}
 ^{(k)}\nabla A^{\mu}\: =\: {^{(k)}}D A^{\mu}\: +
     \: {^{(k)}}\Gamma^{\mu}_{\nu o} A^{\nu} dx^o\: ,
\end{equation}
     where ${^{(k)}}DA^i = DA^i + {^{(k)}}S^i_{k l}A^k dx^l $
     is the absolute differential of the pseudo-Euclidean space
     $V_3 $ (here ${^{(k)}}S^i_{k l} = S_{k l}{^i}
     - S_l{^i}{_k} - S_k{^i}{_l} $
     and $S_{k l}{^i} $ is the tortion tensor) and
     ${^{(k)}}\Gamma^{\mu}_{\nu o} $ is the connection of the space
     $^kV_4 $.

     4. To obtain a more detailed description, the
     definition of the absolute differential
     $^{(k)}\nabla (\cdots) $ is written in standard form
     \cite{p1}, \cite{p4}
\begin{equation}
\label{e5}
 ^{(k)}\nabla A^{\mu}\: =\: (\partial_{\nu} A^{\mu}\: +
     \: \Gamma^{\mu}_{\omega \nu} A^{\omega}) dx^{\nu}\: ,
\end{equation}
     where $2\Gamma^{\mu}_{\omega \nu} =
     2 {^{(k)}}\Gamma^{\mu}_{\omega \nu} +
     Q^{\mu}_{\omega \nu} $,
     at that $Q^{\mu}_{\omega \nu} =
     {^{(k)}}g^{\mu \gamma} ({^{(k)}}Q_{\omega\nu\gamma} +
     {^{(k)}}Q_{\nu\gamma\omega} -
     {^{(k)}}Q_{\gamma\omega\nu}) $ and
     ${^{(k)}}Q_{\mu\nu\omega} =
     - {^{(k)}}\nabla_{\mu} \Big({^{(k)}}g_{\nu \omega} \Big) $;
     ${^{(k)}}\Gamma^{\mu}_{\omega \nu} =
     \Big\{^\mu_{\omega\nu} \Big\} +
     {^{(k)}}S^{\mu}_{\omega\nu} $,
     where $\Big\{^\mu_{\omega\nu} \Big\} $ is the Christoffel symbol
     and ${^{(k)}}S^{\mu}_{\omega\nu} = S_{\omega\nu}{^{\mu}}
     - S_{\nu}{^{\mu}}{_{\omega}} - S_{\omega}{^{\mu}}{_{\nu}} $
     at that $S_{\omega\nu}{^{\mu}} $ is the tortion tensor.

     If it is required that the definition in Eq.(\ref{e5})
     coincide with chat in Eq.(\ref{e4}), the result obtained is
\begin{equation}
\label{e6}
 2{^{(k)}}\Gamma^i_{o j} dx^j\: +\:
    Q^i_{o \omega}dx^{\omega}\: =\:
    Q^i_{j \omega}dx^{\omega}\: =\:
 2{^{(k)}}\Gamma^o_{\mu j} dx^j\: +\:
    Q^o_{\mu \omega} dx^{\omega}\: =\: 0\: ,
\end{equation}
     which must be satisfied if the imbedding described in
     Secs.2 and 3 is to be possible.

     It may readily be demonstrated that the absolute
     differential $^{(k)}\nabla (\cdots) $ of space $^kV_4 $
     defined by Eq.(\ref{e4}) describes a nonmetric transfer
     in $^kV_4 $. In fact
\begin{equation}
\label{e7}
    {^{(k)}}Q_{o o o}\: =\:
   2{^{(k)}}g_{o o}{^{(k)}}S^o_{o o}\: ,\; \;
    {^{(k)}}Q_{i o o}\: =\: - \partial_i {^{(k)}}g_{o o}\: .
\end{equation}
     The remaining ${^{(k)}}Q_{\mu\nu\omega} = 0 $. As a result
     equations (\ref{e6}) takes the form
\begin{eqnarray}
\label{e8}
 {^{(k)}}S^o_{o j} dx^j & = & - \Big\{^o_{o j}\Big\} dx^j -
 2{^{(k)}}S^o_{o o} dx^o\: ,\nonumber \\
\label{e8-1}
 {^{(k)}}S^o_{i j} dx^j & = & -  \Big\{^o_{i j}\Big\} dx^j +
 \Big\{^o_{i o}\Big\} dx^o\: ,\\
\label{e8-2}
 {^{(k)}}S^i_{o j} dx^j & = & -  \Big\{^i_{o j}\Big\} dx^j +
 \Big\{^i_{o o}\Big\} dx^o\: ,\nonumber
\end{eqnarray}
     Hence it is clear that the tortion $S_{\omega\nu}{^{\mu}} $
     is nonzero.

     Thus, the imbedding described in Sec.2 generates in
     $^kV_4 $ a geometry with tortion and a nonzero
     covariant derivative of the metric tensor.

     5. The test particle motion equations
     are now considered. It is desirable for these
     equations to coincide with the geodesic equations
     in $^kV_4 $. Then these equations should take the form
\begin{equation}
\label{e9}
D p^{\mu}\: =\: - ^{(k)}\Gamma^{\mu}_{\nu o} p^o dx^{\nu}\: ,
 \; \; (p^{\mu}\: =\: m dx^{\mu}/d\tau)\: .
\end{equation}
     Taking this into account, the condition
     $dx_i dp^i = dx^o dp^o $ leads to the equation
\begin{equation}
\label{e10}
 {^{(k)}}S^j_{\nu o} dx^{\nu} dx_j\: =\:
 \Big(\Big\{^o_{\nu o}\Big\} + {^{(k)}}S^o_{\nu o}\Big)
 dx^{\nu} dx^o - \Big\{^j_{\nu o}\Big\} dx^{\nu} dx_j\: ,
\end{equation}
     which, together with Eq.(\ref{e8}), describes all the
     nonzero components of ${^{(k)}}S^{\mu}_{\omega\nu} $.

     Further, it is readily evident that, if the components
     ${^{(k)}}S^o_{\nu o} $ are chosen in the form
\begin{equation}
\label{e11}
 {^{(k)}}S^o_{\nu o}\: =\:
 [\partial_{\nu} \ln ((1 - {^{(k)}}g_{o o})/{^{(k)}}g_{o o})]/2\: ,
\end{equation}
     the four-momentum $p^o $ component is found to be
\begin{equation}
\label{e12}
 p^o\: =\: C_1(1 - {^{(k)}}g_{o o})^{-1/2}\: ,
\end{equation}
     where $C_1 = const $. Assuming that $C_1 = m c $, it is found
     that $d\tau = (1 - {^{(k)}}g_{o o})^{-1/2} dt $.

     Hence, Eqs.(\ref{e10}) and (\ref{e11}) are the necessary and
     sufficient conditions for the motion equations Eq.(\ref{e9})
     to be noncontradictory.

     Thus, the classical particle trajectories in the
     potential fields specified with respect to a definite
     IF may be represented as geodesic lines lying on
     isotropic surfaces of some configurational space
     $^kV_4 $ the connection of which has tortion, while
     the transference is nonmetric. The geometry of the space
     $^kV_4 $ has the distinctive property that the magnitude of
     the nonmetricity of the transfer and the tortion are
     determined by specifying the metric coefficient
     ${^{(k)}}g_{o o} $ under the condition that the mixed
     components ${^{(k)}}g_{o i} \equiv 0 $.

     In conclusion, thanks are due to G. I. Flesher for
     interest in the work and for useful critical comments
     and also to I. L. Bukhbinder for fruitful discussions
     of the fundamental aspects of the present work.

\end{document}